\documentstyle[12pt]{article}
\def\be{\begin{equation}}
\def\ee{\end{equation}}

\def\e{\varepsilon}

\def\la{\lambda}

\def\ph{\varphi}

\begin{document}
\renewcommand{\theequation}{\thesection.\arabic{equation}} 
\title{Perihelion advance for orbits with large eccentricities in the Schwarzschild black hole} 
\author{Hans-J\"urgen  Schmidt} 
\date{May 18, 2011}
\maketitle
\centerline{Institut f\"ur Mathematik, Universit\"at Potsdam, Germany} 
\centerline{Am Neuen Palais 10, D-14469 Potsdam,  \  hjschmi@rz.uni-potsdam.de}
\begin{abstract}
We deduce a new formula for the perihelion advance $\Theta$ of a test particle  in the 
Schwarzschild black hole by applying a newly developed non-linear transformation 
within the Schwarzschild space-time. By this transformation we are able to apply 
the well-known formula valid in the  weak-field approximation  near infinity 
 also to trajectories in the strong-field regime near the horizon of the black hole. 
 The resulting formula has the structure $\Theta = c_1 - c_2 \ln(c^2_3 - e^2) $ 
 with positive constants  $c_{1,2,3}$ depending on the angular
 momentum of the test particle. It  is especially useful for orbits with large 
eccentricities $e < c_3 < 1$ showing that $\Theta \to \infty$ as $e \to c_3$. 
\end{abstract}

\vspace{5.mm} 

\noindent 
Keyword(s): Perihelion precession, perihelion advance,  Schwarzschild black hole

\section{Introduction}

Surprisingly little is known about the value of the perihelion advance 
in the strong-field region of the   Schwarzschild space-time, especially
 at large eccentricities of the closed orbit. 
Recent interest in calculations and measurements of orbital characteristics 
 like the relativistic pericenter precession, also called relativistic perihelion 
advance, see e.g.  \cite{k1}, \cite{k2}, \cite{c62} and \cite{k3},  show the necessity to 
have good closed-form expressions for it in the most prominent example 
of a space-time, the   Schwarzschild black hole.  Also the calculation of
perihelion precession in classical mechanics is subject to recent interest,
see \cite{k5} and section 5 of  \cite{s2}. For latest work on analogous questions
in axially symmetric space-times see e.g. \cite{k13}.  In section VII of \cite{e00}, 
the nearly circular motion of  a particle within  $f(R)$-gravity has been discussed, 
 whereas in  \cite{e01}, the orbital motion in  $f(R)$-gravity with 
quadrupole radiation was calculated. 
\par  It is the aim of the present paper to apply the ideas and formulas  deduced in section 
10 of  \cite{s2} to the calculation of the perihelion advance $\Theta$ of periodic orbits 
in the   Schwarzschild black hole.\footnote{Of course, the exact formula is 
well-known:  it contains elliptic integrals,  but in practice this formula is of minor 
use only. And, probably more importantly: these elliptic integrals which can be
evaluated numerically  to every degree of accuracy, do not easily lead to the
 identification of the physically interesting quantities we are trying to find 
out in the present paper.}  Especially, we want to find out, how $\Theta$ depends on 
the eccentricity $e$ of the orbits. To have a stricter posed question, we ask: 
what is the change of  $\Theta$ if we change the eccentricity $e$ of the orbit
 but keep the angular momentum $h = r^2 \dot \ph$ of the particle\footnote{The
dot denotes the derivative with respect to the eigentime of the particle.}
 constant? \par
The core of the deduction is a newly developed non-linear transformation 
within the Schwarzschild space-time. Due to its importance we have chosen to
present this transformation in two independent versions, one by the set of eqs. (2.16)
 to (2.36), which directly applies to the geodesics in the Schwarzschild space-time,
 the other one is given in the appendix, which is a self-contained and more
 abstract deduction of the primarily unexpected symmetry of eq. (2.16).
\par
The notation is as follows:  Let the orbit be the periodic but non-constant 
function $r(\ph)$ with invariantly defined Schwarzschild radius $r$, then $\Theta$ is 
defined by  the period  $2\pi + \Theta$ of the function $r$, i.e. 
$r(\ph) = r(\ph + 2\pi + \Theta)$.  The eccentricity $e$ is defined by
\be
e = \frac{r_2 - r_1}{r_2 + r_1}
\ee
where $r_2 = {\rm max} \,  r(\ph)$ and $r_1 = {\rm min} \,  r(\ph)$.
Thus, $e$ and $\Theta$  are invariantly defined quantities, even for orbits which 
may be very  far from being elliptic ones. Moreover, $e$ does not change if
$r$ is replaced by $c \cdot r$ with a constant $c$. If we identify $e$ and $-e$,
then also the replacement of $r$ by its inverse $1/r$ in eq. (1.1) leads to the same 
eccentricity. \par   By continuously changing $r_1$ and $r_2$ we get circular orbits 
at $r_1 = r_2$, and interchanging  $r_1$ and $r_2$ leads to the 
same geometry of the orbit, hence the same value of   $\Theta$. This means, we 
expect $\Theta$ to be an even function of $e$ at constant angular momentum. 
From now on we restrict $e$ to the interval $0 < e < 1$, nevertheless,  this 
consideration is useful, as the Taylor development of $\Theta$ around $e=0$
should only contain even powers of $e$. \par
 The reason why we parametrize the periodic non-circular orbits by angular
 momentum is the following one: At fixed value $h$, these orbits can be uniquely
be parametrized by $e$, but can also be uniquely be parametrized by 
the perihelion $r_1$ of the orbits, and can also be uniquely parametrized by the total 
energy  $E$ of the test particle. This is not a trivial statement, as for the general case, 
given $h$ and $E$, more than one orbit exists: e.g. one hyperbolic orbit and 
another one leading towards the horizon. 

\section{Geodesics in the Schwarzschild black hole}
\setcounter{equation}{0} 
We take  the Schwarzschild solution in  Schwarzschild coordinates\footnote{But 
see \cite{k4}  for clarifying  historical notes to this notion.}
   with  mass parameter $m>0$ as usual:
\be  
ds^2 = \left(1 - \frac{2m}{r} \right) dt^2 - \frac{dr^2}{1 - 2m/r } - r^2d\Omega^2
\ee
where $d\Omega^2$ is the metric of the standard 2-sphere. We apply units such that light 
velocity $c=1$ and Newton's gravitational constant $G=1$.\footnote{In principle, we 
could also apply units such that $G = 1/m$, and under these circumstances, we have one 
less parameter in all the calculations, but then the departure  from the  usual well-known 
formulas is becoming even larger.} To calculate time-like geodesics we may assume 
without loss of generality that they are situated  in the equatorial plane.
The angular  coordinate is  denoted by $\ph$ and the proper time along the time-like 
geodesic $\left( t(\tau), \, r(\tau), \, \ph(\tau) \right)$  is  denoted by $\tau$,  a dot 
denotes $d/d\tau$. We assume $\dot t > 0$. Then we  get from eq. (2.1)
\be 
1 = \left(1 - \frac{2m}{r} \right) \dot t^2 - \frac{\dot r^2}{1 - 2m/r } - r^2 \dot \ph^2\, .
\ee
We restrict to the region outside the horizon, i.e. to $r > 2m$.
\be
h = r^2 \dot \ph
\ee
 is the conserved angular momentum of the  test particle. We exclude purely radial 
motion which is characterized by $h=0$ and choose the orientation of space such 
that  $h >0$. Inserting  eq. (2.3) into eq. (2.2) we get 
\be 
1 = \left(1- \frac{2m}{r} \right) \dot t^2 - \frac{\dot r^2}{1- 2m/r } - \frac{h^2}{ r^2} \, . 
\ee
A further conserved quantity is the energy $E$ defined by 
\be 
E =  \left(1 - \frac{2m}{r} \right) \dot t  > 0  \, . 
\ee
Inserting eq. (2.5) into eq. (2.4) we get
\be 
1 = \frac{E^2 - \dot r^2}{1 - 2m/r} - \frac{h^2}{ r^2} \, . 
\ee
To remove $\dot r$, the third of the three terms containing a $\tau$-derivative  
from eq. (2.6), we  describe the path of the particle as $r(\ph)$ and get via 
$\frac{dr}{d\ph} = \frac{\dot r}{\dot \ph}$ by the help of eq. (2.3)
\be
\dot r =  \frac{h}{r^2} \cdot \frac{dr}{d\ph}\, .
\ee
Inserting eq. (2.7) into eq. (2.6) we get after  multiplication with $1-2m/r$
\be
\left(1 + \frac{h^2}{r^2} \right) \cdot \left(1 - \frac{2m}{r}\right) = 
E^2 - \frac{h^2}{r^4} \cdot \left(\frac{dr}{d \ph}\right)^2\, .
\ee
To get complete information from  the geodesic equation, we still need 
the radial part of it, we take it from eq. (9.6) of  \cite{s2}: 
\be
0 = \frac{\ddot r}{1-2m/r} - \frac{h^2}{r^3} + \frac{m}{r^2}
 \cdot \frac{E^2 - \dot r^2}{(1 - 2m/r)^2}\, . 
\ee
 Following   \cite{k6}, we introduce the  nondimensionalized  inverted 
Schwarzschild radius  $u$ via
\be
u = \frac{h^2}{m \cdot r}
\ee
and the dimensionless parameter $\e$ via 
\be
\e = \frac{3m^2}{h^2} > 0\, . 
\ee
This leads to $r = h^2/(m \cdot u)$ and
\be
 \frac{dr}{d \ph} = - \frac{h^2}{m \cdot u^2} \cdot \frac{du}{d\ph}\, .
\ee
Inserting eq. (2.12) into  eq. (2.8)  we get  
\be 
\left(1 + \frac{m^2 u^2}{h^2} \right) \cdot \left(1 - \frac{2m^2 u}{h^2}\right) = 
E^2 - \frac{m^2 }{h^2} \cdot \left(\frac{du}{d \ph}\right)^2\, .
\ee
With eq. (2.11)  we finally get
\be 
\left(1 + \frac{\e u^2}{3} \right) \cdot \left(1 - \frac{2 \e u}{3}\right) = 
E^2 - \frac{\e }{3}  \left(\frac{du}{d \ph}\right)^2\, .
\ee
We multiply by $3/(2\e)$ and get 
\be
\frac{1}{2}\left(\frac{du}{d \ph}\right)^2-u+\frac{u^2}{2}-\frac{\e u^3}{3}=\mu
=\frac{3}{2\e} \left(E^2 -1 \right)\, .
\ee
Derivating this equation we get  
\be
\frac{d^2 u}{d\ph^2} + u = 1 + \e u^2\, . 
\ee
Eqs. (2.15)/(2.16) represent the motion of a 
particle $u$ in the
 potential
\be
V(u) =  - u + \frac{u^2}{2} - \frac{\e u^3}{3}
\ee
with $\mu$ interpreted as energy. A dash denoting $\frac{d}{du}$ we get
\be
V'(u) =  - 1 + u - \e u^2
\ee
and
\be
V''(u) =  1 -2 \e u\, . 
\ee

To shift the turning point of this potential to the origin, we define
\be
v = u - \frac{1}{2\e}
\ee
as new variable instead of $u$. Eqs. (2.17)/(2.20) lead to
\be
V(u) = - \frac{1}{2\e} + \frac{1}{12\e^2} - v  + \frac{v}{4\e} - 
\frac{\e v^3}{3} \, .
\ee
Defining further 
\be
\mu_2 = \mu + \frac{1}{2\e} - \frac{1}{12\e^2}
\ee
then eq. (2.15) now reads 
\be
\frac{1}{2}\left(\frac{dv}{d \ph}\right)^2 + v \left(\frac{1}{4\e} -1  \right)
-\frac{\e v^3}{3}=\mu_2 \, . 
\ee
Let us note the special solution $v \equiv 0$ for $\e = 1/4$ and $\mu_2 =0$.  It
 corresponds to a special semistable circular orbit. Apart from this special solution, 
eq. (2.23) possesses periodic solutions $v(\ph)$ only for the parameter range 
$0 < \e < 1/4$. This we will always assume in the following. \par  With the notation 
\be
V_2(v)=  v \left(\frac{1}{4\e} -1  \right) -\frac{\e v^3}{3}
\ee
we get
\be
V'_2(v)=  \frac{1}{4\e} -1   - \e v^2
\ee
and
\be
V''_2(v) =   -2 \e v\, . 
\ee
The equation $V'_2(v)=0$ possesses the solutions 
\be
\pm \frac{\sqrt{1-4\e}}{2\e}
\ee
representing one maximum and one minimum. This forces us to introduce the new 
variable 
\be
w = \frac {2v\e}{\sqrt{1-4\e}}
\ee
instead of $v$. With 
\be
\mu_3 = \frac{4 \e^2 \mu_2 }{1 - 4 \e}
\ee
we get now from 
eq. (2.23) 
\be
\frac{1}{2}\left(\frac{dw}{d \ph}\right)^2 +  \frac{w}{2}\cdot \sqrt{1-4\e}
-\frac{w^3}{6} \cdot \sqrt{1-4\e}=\mu_3 \, . 
\ee
In the next step we replace $\ph$ by a new angular coordinate
\be
\psi = \ph \cdot \sqrt[4]{1-4\e} \, .
\ee
It is to be observed, that $\ph=0$ is identified with $\ph=2 \pi$, so  $\psi=0$ is 
identified with $\psi=2 \pi \cdot  \sqrt[4]{1-4\e}$. Eqs. (2.30)/(2.31) together with 
\be
\mu_4 = \frac{\mu_3}{1-4 \e}
\ee
lead to
\be
\frac{1}{2}\left(\frac{dw}{d \psi}\right)^2 +  \frac{w}{2}-\frac{w^3}{6} =\mu_4 \, . 
\ee
This is the equation we are going to solve now. Derivating eq. (2.33) we arrive at
 the simple equation
$$
\frac{d^2 w}{d\psi^2}  = \frac{w^2 - 1}{2}\, . 
$$
  Remarkably enough, it contains  no $\e$.  The function 
\be
f(w) = \frac{w}{2} -\frac{w^3}{6}
\ee
has zeroes at $w=0$ and $w = \pm \sqrt 3$. For the derivative we get 
\be
f'(w) = \frac{1}{2} -\frac{w^2}{2}
\ee
possessing zeroes at  $w = \pm 1$. $f(-1) = -1/3$ is the local minimum and
$f(1) = 1/3$ is the local maximum of $f$. We note that also $f(-2)=1/3$. The constant
solutions of eq. (2.33) are $w(\psi) \equiv -1$ with $\mu_4 = -1/3$ representing the 
stable circular orbits  and  $w(\psi) \equiv 1$ with $\mu_4 = 1/3$ representing the 
unstable circular orbits.  Besides these exceptions it holds: Every periodic solution 
$w(\psi)$ of eq. (2.33)  is a non-constant one and is completely confined in the interval 
$-2 <w(\psi)< 1$. This is related to the energy parameter $\mu_4$ being confined 
to $-1/3 < \mu_4 < 1/3$. Let $w_1 = {\rm max} \, w(\psi)$ and  
$w_2 = {\rm min} \, w(\psi)$, then $\mu_4 = f(w_1) = f(w_2)$.  
We have $-1 < w_1 < 1$ and $-2 < w_2 < -1$. Let us parametrize these
periodic orbits by the parameter $e_4$ defined by 
\be
w_1 = 2e_4 - 1 \qquad  {\rm  with}  \qquad 0 < e_4 < 1\, .
\ee
In the limit $e_4 \to 0$ we get the stable circular orbits. 
\par
In the other  limit $e_4 \to 1$ we come arbitrarily close to  the unstable circular 
orbits, that means, that the perihelion advance tends to infinity in this limit. \par
Using eqs. (2.34)/(2.36) we get
\be
\mu_4(e_4) = f(w_1)= - \frac{4}{3} e_4^3 + 2 e_4^2 - \frac{1}{3}\, .
\ee
Due to $d\mu_4/de_4 = 4 e_4 (1 - e_4) > 0$ this represents a one-to-one
correspondence between $e_4$ and $\mu_4$. Solving now the equation
$\mu_4 = f(w_2)$ for $w_2$ we get with eq. (2.37)
\be
w_2 = \frac{1}{2}-e_4-\sqrt{3}\cdot \sqrt{1-\left(\frac{1}{2}-e_4\right)^2}\, . 
\ee
Let $\psi_0 = k(e_4)$ be the complete period of the function $w(\psi)$. We calculate 
it by solving eq. (2.33) with $w(0) = w_2$ and $w(\psi_0/2) = w_1$ and get 
\be
\psi_0=k(e_4)=2 \, \int_{w_2}^{w_1} \, \frac{dw}{\sqrt{2\mu_4 + w^3/3 -w}}\, .
\ee
 Though the exact solutions for such integrals  can be found in the literature, see e.g. 
 \cite{k7}, page 355, for the general theory and \cite{k8}, \cite{k11} for its concrete 
application, they are of little use as the  elliptic integrals  can be evaluated by the 
Weierstrass function only, and not in the usual closed-form presentation, which would 
allow for a physical interpretation. The three cases $e_4 \in \{1/2, \, 0, \, 1 \}$ will 
now be considered in detail.  \par 
Let $e_4 = 1/2$, then $\mu_4 = w_1 =0$ and $w_2 = - \sqrt 3$. Eq. (2.39) simplifies to
\be
\psi_0=k(1/2)=2 \, \int_{- \sqrt 3}^{0} \, \frac{dw}{\sqrt{ w^3/3 -w}}\, .
\ee
The substitution $x = \sqrt{-w}/ \sqrt[4]{3}$ leads to 
\be
k(1/2) = 4 \cdot \sqrt[4]{3} \cdot I
\ee
where
\be
I = \int_0^1 \frac{dx}{\sqrt{1 - x^4}} = \frac{1}{4 \sqrt{2\pi}} \cdot
\left( \Gamma(1/4) \right)^2
\ee
according to standard tables,  the  Gamma-function has 
$\Gamma(1/4) = 3.62561$, and  we get $I = 1.31103$ and 
\be
k(1/2) = \psi_0 = 6.90164 = 2 \pi \cdot  1.09843. 
\ee
To get higher accuracy, the $1.09843$ in eq. (2.43) has to be replaced
 by $\alpha$ with
$$
\alpha = \frac{\sqrt[4]{3}}{\sqrt{8 \pi^3}} \cdot \left( \Gamma(1/4) \right)^2 \, .
$$
\par The limit $e_4 \to 0$ corresponds to $w_1 \to -1$, and $f''(-1) = 1$, so the 
system represents a harmonic oscillator with unit frequency:
\be
w(\psi) = -1 + e_4 \cdot \cos \psi\, , \qquad k(0) = \psi_0 = 2\pi\, .
\ee
\par  The limit $e_4 \to 1$ can similarly be solved: it corresponds to $w_1 \to 1$, 
and $f''(1) = - 1$, so we have to replace cos by cosh in eq. (2.44) and $e_4$
by $1 - e_4$: 
\be
w(\psi) =1-(1-e_4) \cdot \cosh \psi\, , \qquad  \psi_0 \to \infty \quad
{\rm as } \quad e_4 \to 1\, .
\ee
To quantify the diverging part we replace $\cosh \psi$ by $\exp(\psi)/2$
and $1 - e_4$ by $(1 - e_4^2)/2$, then eq. (2.45) reads 
$$
w(\psi) = 1 - \exp(\psi )(1 - e_4^2)/4\, .
$$
$w(\psi)=0$ will be reached at $\psi = \ln(4/  \left(1 - e_4^2) \right)$, so the 
general structure must be approximately 
\be
k(e_4) = \psi_0 = c_1 - c_2  \cdot  \ln \left(1 - e_4^2 \right)
\ee
with certain positive constants $c_1$ and $c_2$ of order 1. We will fix them
by the condition that eq. (2.46) holds exactly true for $e_4 =1/2$ and
$e_4 =0$ according to eqs. (2.43) and (2.44) resp. This leads to $c_1 = 2\pi$
and $c_2 = 2\pi \cdot 0.34215$. So eq. (2.46) reads 
\be
k(e_4) = 2\pi \cdot  \left( 1 -  0.34215  \cdot  \ln \left(1 - e_4^2 \right) \right)\, .
\ee
For $e_4 \ll 1$ we get approximately
\be
k(e_4) = 2\pi \cdot  \left( 1 +  0.34215  \cdot   e_4^2  \right)\, .
\ee
To get higher accuracy, the $0.34215$ in eqs. (2.47), (2.48), (2.51) and (2.52) has to be 
replaced  by $\beta$ with $\alpha $ from above  and
$$
\beta = \frac{\alpha - 1}{\ln 4 - \ln 3} \, .
$$
\par
Let us now return from the $\psi$-picture to the $\ph$-picture. Let $\ph_0$
 be the complete period in the $\ph$-picture, then we get by eq. (2.31)
\be
\ph_0 = \frac{\psi_0}{ \sqrt[4]{1-4\e}} =  \frac{k(e_4)}{ \sqrt[4]{1-4\e}}\, .
\ee
The perihelion advance is $\Theta = \ph_0 - 2 \pi$, i.e.
\be
\Theta =   \frac{k(e_4)}{ \sqrt[4]{1-4\e}} - 2 \pi\, .
\ee
Inserting eq. (2.47) into eq. (2.50) we get
\be
\Theta = 2 \pi \cdot 
\left(  \frac{1 -  0.34215  \cdot  \ln \left(1-e_4^2\right)}{ \sqrt[4]{1-4\e}}-1 \right) \, .
\ee
This eq. (2.51) is useful for all $e_4$ with $0<e_4<1$ and a strict result for 
$e_4=1/2$; also both the limiting behaviours $e_4 \to 0$ and $e_4 \to 1$
represent strict ones. \par Inserting the approximation eq. (2.48) into eq. (2.50) we get
\be
\Theta =  2 \pi \cdot 
\left(  \frac{1 +  0.34215  \cdot   e_4^2}{ \sqrt[4]{1-4\e}} - 1  \right) \, .
\ee
This eq. (2.52) is useful for all $e_4$ with $0<e_4 \ll  1$. Inserting the expression
(2.11) for $\e$, we get the final formula, expressing the perihelion advance in dependence
on angular momentum $h$ and the parameter $e_4$, which is correlated to the 
eccentricity of the orbit: 
\be
\Theta = 2 \pi \cdot 
\left(\frac{1-0.34215\cdot \ln \left(1-e_4^2\right)}{\sqrt[4]{1- 12m^2/h^2}}-1\right) \, .
\ee

\section{Properties of the orbits}
\setcounter{equation}{0} 
To be able to interpret eq. (2.53) we need a better knowledge of the orbits.  The formula 
(1.1) for the eccentricity is invariant with respect to  multiplication of $r$ by a constant 
factor, and also (after identification of  $e$ with $-e$)  invariant with respect to a 
replacement  of $r$ by $1/r$, but not invariant with respect to adding a constant to $r$. 
But just this addition of the constant   has been done in eq. (2.20), so it must be expected, 
that the parameter $e_4$  will depend both in $e$ and on $h$, and not only on $e$. \par
Due to eq. (2.11), the condition $2m < r < \infty $ reads $0 < u < 3/(2\e)$, and we restrict
to the interval $0< \e < 1/4$ because  only for  these $\e$-values non-circular periodic 
orbits exist.  The limit $\e \to 0$ is the Newtonian limit; for this case we know the 
non-circular periodic orbits to be exact elliptic ones, i.e. $\Theta =0$ for all values of $h$ 
and $e$. For $\e = 1/4$, only one periodic orbit exists, it is a circular one. For $\e > 1/4$, 
the angular momentum is too small to allow for periodic orbits, the particle goes to 
$r \to \infty$ or to $r \le 2m$  after sufficient long time.  \par   To find the real properties of 
the perihelion advance, we have therefore to restart at the point just before the translation (2.20) 
 has been applied. The function $V(u)$, eq. (2.17), has the following zeroes: $V(0) =0$, for 
 $\e > 3/16$ this is the only one, for $\e \le 3/16$  the other  zeroes of $V$ are calculated via 
\be
u_{5,6} = \frac{3}{4 \e} \cdot \left(1 \pm \sqrt{1 - \frac{16 \e}{3}} \right)
\ee
where $0 < u_6 \le 3/(4 \e) \le u_5 < 3/(2\e)$, and in the special case 
$\e = 3/16$ we have the double root $u_5 = u_6 =  3/(4 \e)$.  \par
Let $u_1$ be the maximal value of $u(\ph)$ and $u_2$ be the minimal one. 
Then we get \footnote{The parameter $p$ has its usual meaning, its relation
 to $r_{1,2}$ and  $e$ can be seen e.g. from eq. (3.9) below.} 
\be
u_{1,2} = \frac{h^2}{m r_{1,2}} = \frac{m(1 \pm e)}{p - m(3 + e^2)}
\ee
leading to 
\be
\frac{u_1 + u_2}{2} = \frac{m}{p - m(3 + e^2)}
\ee
and
\be
\frac{u_1 - u_2}{2} = \frac{m e}{p - m(3 + e^2)}\, . 
\ee
Solutions with constant value $u$ fulfil $u = 1 + \e u^2$, see eq. (2.18), i.e.
the function $dV(u)/du$  has the  two zeroes:
\be
u_{3,4} =   \frac{1}{2 \e} \cdot \left( 1 \pm \sqrt{1 - 4\e} \right)
\ee
where $0 < u_4 < 1/(2 \e) < u_3 < 1/\e$.  The second derivative of $V(u)$, see
eq. (2.19),  vanishes  for $u = 1/(2\e)$ only. Therefore, at $u =u_4$ we have a 
minimum of $V(u)$, and at $u=u_3$ a maximum. For the special case $\e = 3/16$ we 
have $u_3 = u_5 = u_6 = 4$ and $V(4) =0$. \par    We get 
\be 
V(u_4) = \frac{1}{12 \e^2} \cdot\left( 1 - 6\e - (1-4\e)^{3/2} \right)
\ee
and have obviously  always $V(u_4) <0$. To find the sign of $V(u_3) $ 
more calculations are  necessary. We get 
\be 
V(u_3) = \frac{1}{12 \e^2} \cdot\left( 1 - 6\e + (1-4\e)^{3/2} \right)
\ee
which is positive for $0 < \e < 3/16$ and negative for  $3/16 < \e < 1/4$. Thus, to get 
the set of  non-constant periodic bounded orbits, we have to distinguish 
three cases:\footnote{The second case can be subsumed to the first one by allowing 
$\e \le 3/16$ as well as under the third one by allowing $\e \ge 3/16$. But as the second 
case possesses other special properties  it is simpler to  deal with it in an extra case.} 
First case: For  $0 < \e < 3/16$ such orbits exist for $0 < - \mu < - V(u_4)$ with
$\mu$ taken from eq. (2.15).
Second case: For  $ \e = 3/16$ such orbits exist for $0 < - \mu < 16/27$.
Third case:  For  $3/16 < \e < 1/4$ such orbits exist for $ - V(u_3) < -\mu < - V(u_4)$.
\par   We parametrize the bounded solutions by $r_1$ and $r_2$, where $r_1$ is
the perihelion  and $r_2$ the aphelion. In the present calculations we 
restrict to the parameter values $2m < r_1 < r_2 < \infty$, i.e. to motion 
completely outside the horizon $r = 2m$.  \par We  pose as  additional restriction
 to the possible values of $r_1$ and $r_2$  the property, that a 
bounded orbit with these values really exists. \par
Second case: $ \e = 3/16$ implies $h=4m$. All values $0 < - \mu < 16/27$ 
are possible, leading to $25/27 < E^2 < 1$, with $E$ from eqs. (2.5)/(2.14)
and all values of the eccentricity $0<e<1$ are possible. In the $u$-picture we get: 
All values $0<u_2<4/3$ an all values $4/3<u_1<4$ are possible; and this
also exhausts the set of all possible values. In the $r$-picture we get: 
perihelion $r_1$ has $4m < r_1 <12m$, and aphelion $r_2$ has $r_2 >12m$. 
 In the limit $e \to 0$ we get the perihelion advance
 $\Theta \to 2\pi (\sqrt 2 -1)$ and in the limit $e \to 1$ we get the perihelion advance
 $\Theta \to \infty$. \par
First case: $0 < \e < 3/16$, i.e. $h > 4m$, and all values of the 
eccentricity $0<e<1$ are possible. Physically, this first  case can be interpreted as 
follows:  Further increasing the energy of the particle 
at constant angular momentum would have the consequence that its path
goes to $r \to \infty$, so no periodic orbit appears. This behaviour we know 
already from Euclidean geometry: if the eccentricities of a set of ellipses tend
to $1$, then the result is a parabola.  \par
Third case: $3/16 < \e < 1/4$, i.e. $ 2 \sqrt 3 m< h < 4m$, and all values of the 
eccentricity $0<e<e(\e)$ are possible, where $e(\e) < 1$ is the following expression:
\be
e(\e) = -  \frac{3}{1 - 2/\sqrt{1 - 4 \e}}\, .
\ee
In the limit $\e \to 3/16$ we get, as expected, $e(\e) \to 1$. It is, however, 
quite  surprising, that in the other limit $\e \to 1/4$ we get  $e(\e) \to 0$.
The perihelion advance tends to infinity if the eccentricity tends to $e(\e)$.
For $\e$-values being only slightly below $1/4$, the perihelion advance becomes 
quite large even for extremely small eccentricities. Physically, this third case can be 
interpreted as follows:  Further increasing the energy  of the particle  at constant 
angular momentum would have the consequence that its path goes to $r < 2m$ 
inside the horizon, so no periodic orbit appears. \par  To get an easy  comparison 
with  other deductions we also define the arithmetic  mean  $a = (r_1 + r_2)/2$ 
of $r_1$ and $r_2$ and call it semimajor axis. The geometric mean of them 
is denoted by $ b = \sqrt{r_1 \cdot r_2} $ and we call it semiminor axis. 
 Even if the orbit is  not  an exact ellipse, we use the usual formula (1.1)
 for defining the eccentricity $e = (r_2 - r_1)/(r_2 + r_1)$ having the  
range $0 < e < 1$. Further, we use the parameter $p = b^2/a$ which is also sometimes
 used in  dealing with ellipses, it holds $p = a(1-e^2) = 2r_1 r_2 /(r_1 + r_2)$, so
$p$ is just the harmonic mean of $r_1$ and $r_2$. \par
For the exact ellipse in the Euclidean plane we have the equation 
$$
r = \frac{p}{1 + e \cos \ph}
$$
thus 
\be
r_1 = \frac{p}{1 + e } \qquad {\rm and}  \qquad   r_2 = \frac{p}{1 - e}\, .
\ee
Therefore, the parametrization of  the set of orbits discussed here with
$(r_1, \, r_2)$ can also be done with $(p, \, e)$, and the latter one eases the 
comparison  with the literature. \par
 In the next step we prescribe the 
values of $r_1$ and $r_2$ and calculate the 
energy $E$ and the angular 
momentum $h > 0$ by use of eq. (2.8). 
 At the extrema, $dr/d\ph =0$, so we get as one of the conditions 
\be
\left(1 + \frac{h^2}{r_1^2} \right) \cdot \left(1 - \frac{2m}{r_1}\right) = 
\left(1 + \frac{h^2}{r_2^2} \right) \cdot \left(1 - \frac{2m}{r_2}\right) \, .
\ee
Inserting eqs. (3.9) into eq. (3.10) we arrive at 
\be
mp^2 = h^2 \left( p - m(3 + e^2) \right)
\ee
which makes sense for $p > m(3 + e^2)$ only. This inequality will always 
 be assumed to hold in the following, it is equivalent to
\be
r_1 > m \cdot \frac{3 + e^2}{1+e}\, ,
\ee
meaning that the minimally allowed value for $r_1$ depends on the eccentricity,  
it  holds 
\be
2m < m \cdot \frac{3 + e^2}{1+e} < 3m\, .
\ee
Analogously we get 
\be 
r_2 > m \cdot \frac{3 + e^2}{1-e} > 3m\, .
\ee
Inequality (3.12) can also be expressed as $ab^2 > m(4a^2 - b^2)$. From eq. (3.11) 
we get
\be
h = \frac{p \sqrt m}{\sqrt{p - m(3 + e^2)}}
\ee
and  then
\be
E = \frac{\sqrt{(p-2m)^2 - 4m^2e^2}}{\sqrt{p( p - m(3 + e^2))}} \, .
\ee
Further, the parameter $\e$ calculates to
\be
\e = \frac{3m^2}{h^2} = \frac{3m}{p} - \frac{3m^2(3+e^2)}{p^2} \, .
\ee
One easily calculates 
\be
0 < \e \le \frac{3}{4(3+e^2)} < \frac{1}{4}
\ee
and for $p=2m(3+e^2)$  one has equality  in the middle of this chain of inequalities.

\section{Discussion}
\setcounter{equation}{0} 
Let us interpret eq. (2.53), i.e.
\be
\Theta = 2 \pi \cdot 
\left(\frac{1-0.34215\cdot \ln \left(1-e_4^2\right)}{\sqrt[4]{1- 12m^2/h^2}}-1\right) \, .
\ee
This fourth root in the denominator seems at least a little bit dubious, 
so we compare with  the literature. First we concentrate on the orbits with negligible 
eccentricity but allow strong fields, i.e., the orbit may be close to the horizon. 
Then  eq. (4.1) reduces to 
\be
\Theta = 2 \pi \cdot 
\left(\frac{1}{\sqrt[4]{1- 12m^2/h^2}}-1\right)
= 2 \pi \cdot 
\left(\frac{1}{\sqrt[4]{1- 4 \e}}-1\right) \, .
\ee
Second, we apply  $1/\sqrt[4]{1-\delta} = 1 + \delta/4 + 5 \delta^2/32 + \dots$
and get the weak-field limit by inclusion of the first two terms to
\be
\Theta = 2 \pi \cdot \left( \frac{3m^2}{h^2}+ \frac{45m^4}{2h^4}  \right) \, .
\ee
 Eq. (4.40) from \cite{k11} reads in our notation
\be
\Theta = 2\pi \cdot \left(\frac{3m^2}{h^2} + \frac{15 m^4(6 + e^2)}{4h^4} \right)\, .
\ee
As one can see, eqs. (4.3) and (4.4) become identical in the limit $e \to 0$, thus 
confirming the correctness of our calculation at least in this order of approximation. 
\par
To have a better comparison with other known results, we continue the discussion with 
the exact circular orbits. In our  notation they can be calculated by inserting $\dot r =0$
 into eqs. (2.6) and (2.9), i.e.,
\be
 \left( 1 + \frac{h^2}{ r^2} \right) \cdot \left(1 - \frac{2m}{r} \right)  = E^2
\ee
and
\be
\frac{h^2}{r^3}  = \frac{m E^2}{ r^2 (1 - 2m/r)^2}
\ee
resp. These equations can be solved for $h$ and $E$ by
\be
h =\frac{\sqrt{ mr}}{\sqrt{ 1 - 3m/r}} \quad  {\rm and} \quad
E = \frac{1 - 2m/r}{\sqrt{1-3m/r}} \, .
\ee
Therefore, only for $r>3m$ such orbits are possible, and in the limit 
$r \to 3m$, a light-like circular orbit appears: The velocity of the particle is
\be
\frac{\sqrt{ m/r}}{\sqrt{ 1 - 2m/r}}
\ee
which tends to $1$ as $r \to 3m$. For every $r > 3m$, a circular orbit exists, 
and with $h$ from eq. (4.7) we calculate the parameter $\e = 3m^2/h^2$ to
\be
\e = \frac{3m}{r} \cdot \left( 1 -  \frac{3m}{r} \right)\, .
\ee
If we insert the expression $\e$ eq. (4.9) from the circular
 orbits into 
eq. (4.2) we get 
\be
\Theta = 2\pi \cdot \left(\frac{1}{\sqrt[4]{1-12m(1-3m/r)/r} }-1\right) \, .
\ee
This is exactly the same as
\be
\Theta = 2\pi \cdot \left(\frac{1}{\sqrt{1-6m/r} }-1\right)
\ee
which represents the expression for orbits close to circular ones already deduced in 
\cite{s2}, eq. (11.2). For $m \ll r$ we  develop eq. (4.11)   to 
\be
\Theta = \frac{6 \pi m}{r} = \frac{6 \pi m}{a(1-e^2)}
\ee
a form which can be found in the majority of texts, e.g. as eq. (1) of  \cite{k8}. 
In the weak-field approximation applied to eq. (4.8) we get
 velocity $\sqrt{m/r}$, and for the unit mass test particle $E_{\rm kin} = m/(2r)$, 
$E_{\rm pot} = - m/r$, hence total energy $E =1 - m/(2r)$. \par
Let us finally discuss the case $e_4 = 1/2$ in more details. Then the 
perihelion advance reads exactly also for strong fields, see eqs. (2.43)/(2.53):
\be
\Theta = 2 \pi \cdot 
\left(\frac{1.09843}{\sqrt[4]{1- 12m^2/h^2}}-1\right) \, .
\ee
We want to calculate the properties of the orbits to which this case belongs, 
 the eccentricity should be something like $e=1/2$, but, as already said, the 
relation between $e$ and $e_4$ may slightly depend on $h$. Looking back to 
eq. (2.40) we see that $\mu_4 = 0$ and $w_1 = {\rm max} \, w =0$
and $w_2 = {\rm min} \, w = - \sqrt 3$. With $\e = 3m^2/h^2$ and eqs. (2.29)/(2.32)
we have $\mu_2 = \mu_3 =0$. With eq. (2.28) we then get:
$$
v_1 = {\rm max} \, v =0 \qquad {\rm  and} \qquad 
v_2 = {\rm min} \, v = - \sqrt{3 - 12 \e}/{2\e} \, . 
$$
With eq. (2.20) we then get:
$$
u_1 = {\rm max} \, u = \frac{1}{2\e} \qquad {\rm  and} \qquad 
u_2 = {\rm min} \, u = \frac{1}{2\e} \cdot \left(1 - \sqrt{3 - 12 \e}\right) \, . 
$$
To represent a closed orbit,
 $u_2 > 0$ is
 necessary, i.e. $1/6 < \e < 1/4$. 
Applying $r= h^2/(mu)$ leads to the perihelion 
\be
r_1 = 6m
\ee
and aphelion
\be
r_2 = 6m/\left(1 - \sqrt{3 - 12 \e}\right) \, . 
\ee
This gives rise to the eccentricity
\be
e = \frac{1}{2/  \sqrt{3 - 12 \e}   -1 } 
\ee
thus having $e \to 0$  and $\Theta \to \infty$ as $\e \to 1/4$; and $e \to 1$  and 
 $\Theta \to 2 \pi \cdot 0.4456$ as $\e \to 1/6$. This means: For all periodic orbits 
with perihelion $r=6m$, eq. (4.13) represents an exact formula for the perihelion advance.  
\par To find out, how the perihelion advance changes with eccentricity, we solve
eq. (4.16) for $\e$:
$$
\e = \frac{1}{4} - \frac{1}{3} \cdot \frac{e^2}{(e+1)^2}\, . 
$$
Then eq. (4.13) reads
\be
\Theta = 2 \pi \cdot 
\left(\frac{\sqrt[4]{3}}{\sqrt 2}\cdot 1.09843 \cdot \sqrt{1 + 1/e} -1\right) \, ,
\ee
where $1.09843$ is an abbreviation for
$$
 \frac{\sqrt[4]{3}}{\sqrt{8 \pi^3}} \cdot \left( \Gamma(1/4) \right)^2 \, .
$$
This shows that at least in this range, a Taylor development around $e=0$
 is not possible. \par
Next, we want to find out, how the perihelion advance changes with
aphelion
\be
r_2 = 6m \cdot \frac{1 + e}{1 - e}
\ee
leading to 
\be
\Theta = 2 \pi \cdot 
\left(\sqrt[4]{3}\cdot 1.09843 /  \sqrt{1 - 6m/r_2} -1\right) \, .
\ee
The formulas deduced here allow to evaluate orbital properties, especially the 
perihelion advance for {\it  all} periodic orbits in the Schwarzschild field,
 and in regions near   the horizon they are much better and easier to handle
than other methods used in the literature.  
 The method used has the potential also to be applied to alternative 
theories of gravitation, see e.g. \cite{sc136} and  \cite{sc139} and the
references cited there, to get the theoretical background for possible 
experimental tests of the theories.

\section{Appendix}
\setcounter{equation}{0} 
The main idea of this paper is outlined by the set of transformations between
eqs. (2.16) and (2.36). To find out the internal structure of this idea, we present
in this appendix another and independent deduction of the unexpected non-linear 
transformation of eq. (2.16), i.e.
\be
\frac{d^2 u}{d\ph^2} + u = 1 + \e u^2\, ,
\ee
which was the core of the calculations in the main part of this paper.  
It is a self-contained deduction, so it can be read independently from the main 
text, and shows the internal symmetry of  eq. (2.16)/(5.1). \par
Here we concentrate solely on this eq. (5.1) and how  solutions at different
values  $\e$  are related to each other. 
Here, $\e < 1/4$ is a positive real parameter  and $u = u(\ph)$ is a non-constant 
periodic  function of the independent variable   $\ph$. Let the smallest period
 of $u$ be $\vartheta >0$. Due to the simple structure 
of the equation one can easily show that $\vartheta$ can also be defined as 
follows: If  $u(\ph_1)$ is  a local minimum  and $u(\ph_2)$ is the next one, 
then $u(\ph_1)= u(\ph_2)$  and $\vartheta = \ph_2 - \ph_1$. Thus, the perihelion 
shift  for this orbit is
\be
\Theta =\vartheta - 2 \pi  \, .
\ee
Now we fix a further real  parameter $\lambda >1$ and define the  function 
$v = v(\ph) $   as follows: 
\be
v(\ph) = B \cdot u( \sqrt{\lambda}  \, \ph) - A
\ee
where $A$ and $B$ depend on $\e$ and $\la$ only, and $B > 0$. Clearly, 
$v(\ph)$ is also a non-constant periodic function.
The function $v$ has smallest period $\tilde \vartheta =   \vartheta /  \sqrt{\la} $
 and   perihelion shift $\tilde \Theta = \tilde \vartheta - 2 \pi$. With eq. (5.2)  we get
\be
\tilde \Theta =   \Theta / \sqrt{\la} - 2 \pi (1 - 1/\sqrt{\la})\, . 
\ee
Thus, the perihelion shift of the function $v$ is smaller than simply the 
perihelion  shift of the function $u$ divided by $\sqrt{\la}$ as one would
 have expected from a first glance. In this sense the 
transformation discussed here is a non-linear one.  
Now we want to fix    $A$ and $B$ such that 
\be
\frac{d^2 v}{d\ph^2} + v = 1 + \tilde \e v^2\, 
\ee
with a parameter  $ \tilde \e  = \tilde \e (\e, \la)$. This means, that $v$ shall
solve essentially the same equation as $u$ does, only the value of the 
parameter $\e$ may differ. 
 To find the form of $A$, $B$
and $\tilde \e$ we insert (5.1) 
and (5.3) into (5.5). This leads 
via
$$
\frac{d^2 v}{d\ph^2}  =  b \cdot \frac{d^2 u(\sqrt{\lambda}  \, \ph)}{d\ph^2}
$$
to
\be
B\la [1 + \e u^2 -u] + B u(\sqrt \la \ph) -A=
1 + \tilde \e[B^2 u^2+ A^2 - 2AB u] \, .
\ee
Here, $u$ means $u(\sqrt \la \ph)$. 
This equation (5.6) must be identically fulfilled. The vanishing of the terms proportional 
to $u^2$ leads to $B \la \e = B^2 \tilde \e $, i.e. to
\be
\tilde \e = \frac{\la \e}{B}\, .
\ee
Inserting (5.7) into (5.6) we get
\be
B\la - B \la u(\sqrt \la \ph) + B u(\sqrt \la \ph) - A = 1 + A^2 \la \e /B
- 2A \la \e u(\sqrt \la \ph)\, . 
\ee
The vanishing of the terms proportional to $u$ leads to 
$$
- B \la + B = - 2 A \la \e \, , 
$$
i.e. to
\be
A = \frac{B (\la - 1)}{2 \la \e}  \, . 
\ee
One gets $A > 0$. Finally, inserting (5.9) into (5.8) we get 
$$
B \la - \frac{B (\la - 1)}{2 \la \e} = 1 + \frac{\la \e}{B} \cdot
\left(    \frac{B (\la - 1)}{2 \la \e}  \right)^2   \,  , 
$$
i.e.
\be
B = \frac{4 \la \e}{1 - \la^2 (1 - 4 \e)}\, . 
\ee
To ensure $B>0$ we restrict to the region where $\la^2(1-4\e)<1$, i.e. where
\be
1 < \la < \frac{1}{\sqrt{1 - 4\e}}\, . 
\ee
Inserting (5.10) into (5.7) and (5.9) we get
\be
\tilde \e = \frac{1}{4} \cdot [ 1 - \la^2(1-4\e) ]  > 0
\ee
and
\be
A = \frac{2 (\la - 1)}{1 - \la^2(1-4\e)}
\ee
resp. To find the upper limit for $\tilde \e$ we calculate 
\be
\e - \tilde \e = \left(\frac{1}{4} - \e \right) \cdot \left(\la^2 - 1\right)
\ee
which leads to $0 < \tilde \e < \e$. \par 
The limit
$\tilde \e \to 0$ is a singular one, as in this limit, the non-linear equation 
 (5.5) becomes a linear one.  Nevertheless, it leads to the following result: 
All non-constant solutions of eq. (5.5) with $\tilde \e = 0$ have the period
$ \tilde \vartheta = 2 \pi$, i.e. perihelion shift $\tilde \Theta = 0$.
With eq. (5.4) we get for this case
\be
\Theta = 2 \pi \left(\sqrt \lambda - 1\right) \, .
\ee
With eq. (5.12) we get for $\tilde \e = 0$ 
\be
\lambda = (1 - 4 \e)^{-1/2}\, .
\ee
Combining eqs. (5.15) and (5.16) we finally get for the perihelion shift
\be
\Theta = 2 \pi \left( \frac{1}{\sqrt[4]{1 - 4 \e}}
 - 1\right) \, .
\ee
This is exactly the same as eq. (4.2).

\end{document}